\documentclass[sn-basic]{sn-jnl}% Basic Springer Nature Reference Style/Chemistry Reference Style

\usepackage{graphicx}%
\usepackage{multirow}%
\usepackage{amsmath,amssymb,amsfonts}%
\usepackage{amsthm}%
\usepackage{mathrsfs}%
\usepackage[title]{appendix}%
\usepackage{xcolor}%
\usepackage{textcomp}%
\usepackage{manyfoot}%
\usepackage{booktabs}%
\usepackage{algorithm}%
\usepackage{algorithmicx}%
\usepackage{algpseudocode}%
\usepackage{listings}%

\graphicspath{{images/}}
\usepackage{caption}
\usepackage{subcaption}
\usepackage{siunitx}
\usepackage{gensymb}
\usepackage[export]{adjustbox}
\usepackage{wrapfig}
\usepackage{sidecap}
\usepackage{amssymb}
\usepackage{ulem}
\theoremstyle{thmstyleone}%
\newtheorem{theorem}{Theorem}%  meant for continuous numbers
\newtheorem{proposition}[theorem]{Proposition}% 
\theoremstyle{thmstyletwo}%
\newtheorem{example}{Example}%
\newtheorem{remark}{Remark}%
\theoremstyle{thmstylethree}%
\newtheorem{definition}{Definition}%
\begin{document}

\title[Article Title]{Quantifying the performances of SU-8 microfluidic devices: high liquid water tightness, long-term stability, and vacuum compatibility}

\author[1,2]{\fnm{Said} \sur{Pashayev}}         %\email{iauthor@gmail.com}

\author[1,3]{\fnm{Romain} \sur{Lhermerout}}    %\email{iiauthor@gmail.com}

\author[1]{\fnm{Christophe} \sur{Roblin}}      %\email{iiauthor@gmail.com}

\author[1]{\fnm{Eric} \sur{Alibert}}         %\email{iiauthor@gmail.com}

\author[1]{\fnm{Jerome} \sur{Barbat}}      %\email{iiauthor@gmail.com}

\author[1]{\fnm{Rudy} \sur{Desgarceaux}}     %\email{iiauthor@gmail.com}

\author[1]{\fnm{Remi} \sur{Jelinek}}     %\email{iiauthor@gmail.com}

\author[1]{\fnm{Edouard} \sur{Chauveau}}      %\email{iiiauthor@gmail.com}

\author[1]{\fnm{Saïd} \sur{Tahir}}      %\email{iiiauthor@gmail.com}

\author[1]{\fnm{Vincent} \sur{Jourdan}}      %\email{iiiauthor@gmail.com}

\author[2]{\fnm{Rasim} \sur{Jabbarov}}      %\email{iiiauthor@gmail.com}

\author[1]{\fnm{Francois} \sur{Henn}}      %\email{iiiauthor@gmail.com}

\author*[1]{\fnm{Adrien} \sur{Noury}}\email{adrien.noury@umontpellier.fr}

\affil[1]{\orgname{Laboratoire Charles Coulomb (L2C), Univ. Montpellier, CNRS},  \city{Montpellier}, \country{France}}    
%\state{State},\orgdiv{Department},\orgaddress{\street{Street}, \postcode{100190},
\affil[2]{\orgname{Institute of Physics of Azerbaijan National Academy of Sciences}, \city{Baku}, \country{Azerbaijan}} %\state{State},
\affil[3]{\orgname{Laboratoire Interdisciplinaire de Physique, Univ. Grenoble Alpes \& CNRS}, \city{Grenoble}, \country{France}}  %\state{State},

\abstract{Despite several decades of development, microfluidics lacks a sealing material that can be readily fabricated, leak-tight under high liquid water pressure, stable over a long time, and vacuum compatible. In this paper, we report the performances of a micro-scale processable sealing material for nanofluidic/microfluidics chip fabrication, which enables us to achieve all these requirements. We observed that micrometric walls made of SU-8 photoresist, whose thickness can be as low as 35 \textmu m, exhibit water pressure leak-tightness from 1.5 bar up to 5.5 bar, no water porosity even after 2 months of aging, and are able to sustain under \unboldmath$10^{-5}$ mbar vacuum. This sealing material is therefore reliable and versatile for building microchips, part of which must be isolated from liquid water under pressure or vacuum. Moreover, the fabrication process we propose does not require the use of aggressive chemicals or high-temperature or high-energy plasma treatment. It thus opens a new perspective to seal microchips where delicate surfaces such as nanomaterials are present.}

%\keywords{SU-8, microfluidics, high burst pressure, ionic leakage, vacuum compatibility, ageing}

\maketitle

\section{Introduction}
Sealing is one of the inevitable processes in micro and nanofluidic chip fabrication. Ideally, low-temperature sealing methods are desirable to use with sensitive materials and avoid the thermal expansion of substrates. Also important is to achieve elevated burst pressure tightness, both for the reliability of the device and to operate small channels with large fluidic resistance. Additionally, microfluidics is gaining popularity for highly sensitive mass measurements in liquid environments to weigh biomolecules, single cells, etc. In this measurement, experiments are carried out in a vacuum with microfluidic nanomechanical resonators \cite{8808536,Burg2007}. Therefore, low-temperature bonding, high liquid water pressure leak-tight, vacuum compatible, microscale processable material is requested.\\
Various methods are used in microfluidics such as plasma bonding \cite{C3LC41345D}, thermal fusion \cite{Tsao2009BondingOT}, and anodic bonding \cite{GRAY199957}, in order to seal conformal materials like PDMS, non-conformal materials like plastics, or hard surfaces (e.g. glass, silicon wafers). These materials and methods typically require at least one of the following conditions: activation of the extremely clean surfaces, and surface chemistry \cite{TEMIZ2015156,Liu2021AUB}. So far, PDMS has been the most common material in chip sealing. However, this material has several drawbacks such as the adsorption of hydrophobic molecules, poor stability after surface treatment, swelling by organic solvents, vapor water permeability, and breaking under high-pressure operations \cite{doi:10.1021/ac071903e,C1LC20514E}. Moreover, sealing with PDMS requires activating at least one of the surfaces, typically using O\textsubscript{2} plasma. This can limit its application because surface activation can damage any delicate material used on the surfaces, such as nanomaterial \cite{MATHUR2012425}.\\
Photoresists such as SU-8, when used as an adhesive material to bond hard surfaces, might present better performances than PDMS while allowing bonding substrates without surface activation or high temperature. Indeed, SU-8 is an epoxy-based negative tone photoresist, developed by IBM in the mid-1990s, and initially used as an inexpensive mold maker. Soon after it was used for the fabrication of high aspect ratio MEMS \cite{Lorenz1997SU8AL} and later became a widely used material in microfluidics, particularly for replica molding PDMS or other polymers \cite{Kamande2015CloningSS}. Interestingly, SU-8 being photosensitive allows forming of patterns by photolithography with micrometric dimensions, very well suited for microfluidics. SU-8 is also used as both a structural material and an intermediate layer for adhesive bonding \cite{Salvo2012AdhesiveBB}, exhibiting excellent physical and optical properties, good chemical stability, and low porosity \cite{Narayan_2018}. However, its leak-tightness while it is submitted to a liquid under pressure, stability, and vacuum compatibility is still to be quantified. Furthermore, SU-8 can be subject to outgassing, as reported from mass spectrometry and gas chromatography techniques \cite{MELAI2009761}. This clearly indicates that a hard bake is needed to boost the cross-linking degree of the resin, to remove the remaining initial constituents, and hence to provide the best performance when it has to be used under vacuum.\\
Bonding strength measurements are mainly based on three methods: tensile strength \cite{5f265656705d4d31837fe585e1934fcc}, burst pressure \cite{2021}, and liquid leakage \cite{doi:10.1063/1.5086611}. The bonding devices where SU-8 was used as an adhesive material had been mainly tested with the tensile strength method. It was reported that the bonding strength was in the range of from 10 bar \cite{Admassu2021} up to 450 bar \cite{Steigert_2008}. However, the tensile strength method does not provide practical information on device performance in use conditions. The most practical method to quantify the strength of the bonding and assess the functionality of a microfluidic chip is the liquid burst pressure test. The liquid burst pressure tightness of the SU-8-based devices when it is used as an adhesive material has been reported as ranging from 1 bar \cite{SipFolch2010} when bonding PDMS layers up to the highest 38 bar \cite{Lima2013} when bonding glass substrates. The durability of the so far reported SU-8-based device was 1-2 days \cite{Narayan_2018}. In comparison, the performances of the PDMS-based devices were reported with the tensile strength from 3.2 bar \cite{Bhattacharya2005} up to 20 bar \cite{Hammami2022}, with the liquid burst pressure method from $\sim$2 bar \cite{GonzalezGallardo2021} up to the highest 8.5 bar \cite{Hammami2022}, when adhesion is enhanced by plasma pre-treatment of the surfaces and the durability from 3 days \cite{Song2018} or the longest of 1 month \cite{BARAKET2013332}.\\

To the best of our knowledge, the quantitative analysis of SU-8 microfluidic device durability, ionic leakage, and water vapor leakage under vacuum has not been reported.\\ 

In this work, we propose an easy and permanent sealing process to quantify the bonding performance of the microfluid device when SU-8 is used as an adhesive material to bond glass with silicon/silicon dioxide or quartz. The sample fabrication and bonding require basic clean room facilities and low temperatures. Since our goal is to develop the sealing process of microchips, our investigation focuses on the search for the thinnest possible wall of SU-8 that can exhibit high-pressure tightness, high vacuum compatibility, and no liquid water porosity over a long time. The latter criterion is checked using an original and very sensitive method based on ionic conduction measurements. The sealing method investigated here is non-destructive, reliable, versatile, and does not require the use of aggressive chemicals, high temperatures, or high-energy plasma treatment. It provides up to 5.5 bar liquid water burst pressure leak-tightness, vacuum compatibility, and extremely low liquid porosity over long times. 

\section{Materials and Methods}
\subsection{SU-8 patterning.}   
SU-8 preparation is performed in the cleanroom. Because of its improved adhesion to glass (as specified by the manufacturer i.e., Kayaku - MicroChem®), we employed SU-8-3025 instead of the more widely used SU-8-20xx series. Optimal adhesion is obtained using Omnicoat (Kayaku - MicroChem®) as a primer. A 4 mm thick glass substrate is used to fabricate SU-8 patterns. First, it is milled to allow the subsequent placement of capillaries of 1.6 mm diameter for the fluidic connection (Figure \ref{fig:reservoir}). In the case of ionic conductivity measurements, two holes are additionally drilled to insert Ag/AgCl electrodes (Figure \ref{fig:two reservoir }). Surface cleanliness of the substrate is ensured by Piranha cleaning (3:1 ratio H\textsubscript{2}SO\textsubscript{4}:H\textsubscript{2}O\textsubscript{2}). Then, Omnicoat primer is spin-coated in 2 steps: first at 500 rpm for 5 s and then at 3000 rpm for 30 s and baked at 200 °C for 2.5 minutes. The substrate is cooled down to 70 °C and a few milliliters of SU-8 are poured all over the substrate. This temperature is recommended in order to remove bubbles \cite{WinNT}. Once the reaction between the Omnicoat primer and the SU-8 is complete, the substrate is cooled down to room temperature and spin-coated in 2 steps: first at 500 rpm for 5 s and then at 2000 rpm for 30 s. The SU-8 coated substrate is then baked at 100 °C for 35 minutes and slowly cooled down to room temperature. The coating is exposed for 45 s to 365 nm UV photolithography, with a power density of 5.8 \unit{\mW\per\square\cm} through a photolithography mask. A second baking treatment at two different temperatures, i.e., 70° C during 2.5 min and 100 °C during 6.5 min, is applied, followed by slowly cooling down to room temperature. Finally, the pattern of SU-8 is developed for 15 minutes in SU-8 Developer (Kayaku - MicroChem®) and rinsed with isopropanol. The height of the so-obtained SU-8 pattern is 35 \textmu m.

%mW/cm\textsuperscript{2}
\subsection{Device assembling, Bonding \& Fluidic connections}

The SU-8 pattern on the glass substrate is bonded on a Si/SiO\textsubscript{2} wafer to form the reservoirs to be tested. We used Si wafer with 90 nm SiO\textsubscript{2} dry oxide. The Si wafer was not treated with Omnicoat or any adhesion promotor. 2 mm thick brass rectangular frames are placed on either side and maintained with 4 screws (Figure \ref{fig:reservoir}c). 
The influence of pressure applied on the SU-8 pattern is assessed by observing optically the device (Figure S1). This crucial step is strictly controlled to obtain homogeneity by observing interferences/reflectivity with an optical microscope in order to ensure the adhesion of the SU-8 wall on both the glass substrate and the wafer and to avoid any deformation of the pattern. The last step of the process consists of thermal treatment at 200 °C for 2 hours. It allows us to obtain surfaces that are completely in contact with the SU-8 pattern (Figure S1c). Noteworthy, during this baking step the thickness of the SU-8 layer extends compare to its fabrication thickness from 20 \textmu m to 35 \textmu m, 30 \textmu m to 50 \textmu m, 60 \textmu m to 85 \textmu m, and 100 \textmu m to 135 \textmu m. Note that the frames are kept after bonding and during all experiments reported here. After assembly, a capillary is inserted into the milling area of the glass substrate and sealed with Stycast 2850FT epoxy glue. The device is left at 65 °C for 2 hours in order to let the Stycast ensure efficient sealing. 

%The thickness of the so fabricated SU-8 pattern is typically 20 \textmu m, which expanded during the last step of assembling process up to 35 \textmu m.
%%%%%%%%%%%%%%%%%%%%%%%%%%%%%%%%%%%%%
\section{Results and discussion}
\subsection{Liquid water pressure tightness}

\begin{figure}[h]
    \centering
    \begin{subfigure}[c]{0.6\textwidth}
        \centering
        \begin{minipage}[t]{\textwidth}
            \vspace{0pt}
            \textbf{(a)}
        \end{minipage}
        \includegraphics[width=\textwidth]{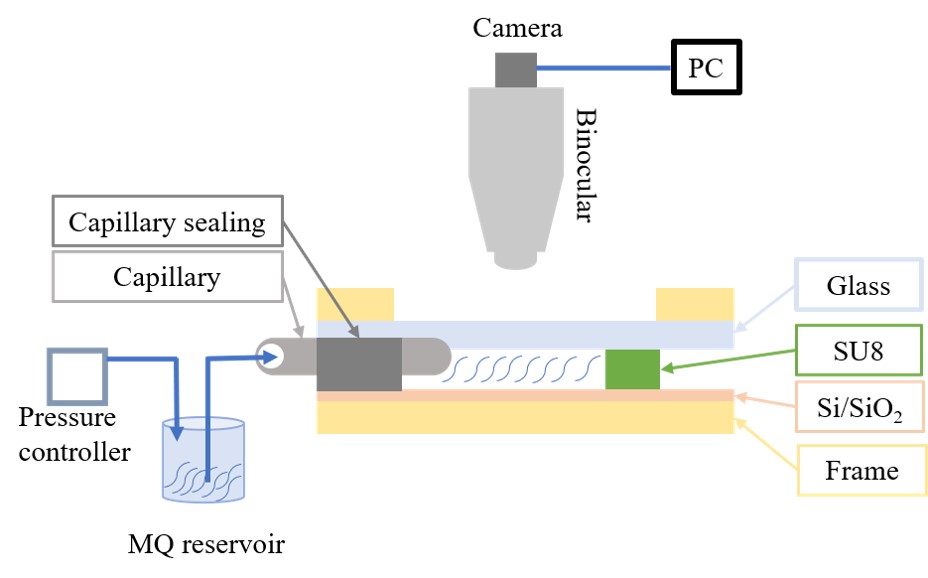}
    \end{subfigure}
 %   \hfill
    \begin{minipage}[c]{0.35\textwidth}
        \begin{subfigure}[b]{\textwidth}
            \centering
            \begin{minipage}[t]{\textwidth}
                \vspace{0pt}
                \textbf{(b)}
            \end{minipage}
            \includegraphics[width=\textwidth]{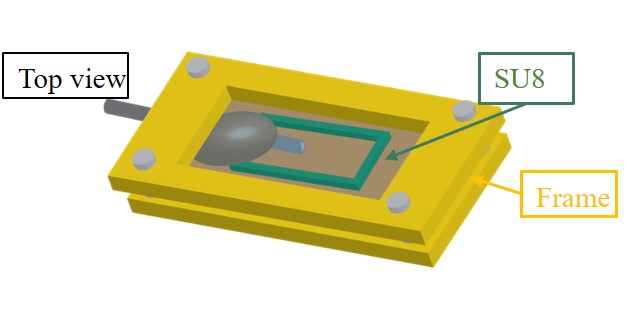}
        \end{subfigure}
        \vspace{1em}
        \begin{subfigure}[b]{\textwidth}
            \centering
            \begin{minipage}[t]{\textwidth}
                \vspace{0pt}
                \textbf{(c)}
            \end{minipage}
            \includegraphics[width=\textwidth]{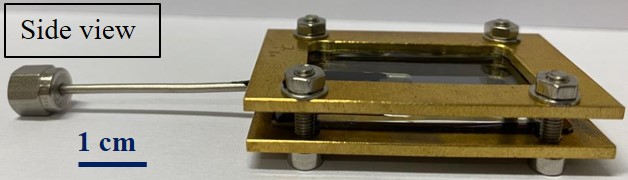}
        \end{subfigure}
    \end{minipage}
  \captionsetup{font={small, it}}
  \caption{Liquid water pressure leak-tightness measurement setup. (a) Schematic of the setup. A pressure controller allows applying water pressure to the SU-8 reservoir. A binocular microscope is used for visual observation of the material's performance. (b) 3D artistic view of the device. (c) Photography of the device with frames added to improve bonding of the material, and capillary for fluidic connection to the pressure controller.}
  \label{fig:reservoir}
\end{figure}

Water pressure leak-tightness is visually assessed with an optical microscope, change in the contrast/reflectivity indicates the presence of water outside of the SU-8 reservoir. A schematic of the setup is shown in Figure \ref{fig:reservoir}a. Our process allows us to build a SU-8 reservoir whose wall thickness is as small as 35 \textmu m. The pattern is designed on the glass substrate in order to form a microfluidic reservoir. This reservoir is filled with water at a pressure piloted by an Elveflow® OB1 MK3+ pressure controller. We tested SU-8 wall thicknesses ranging from 135 \textmu m down to 35 \textmu m. The given two boundary dimensions are set by our method of chip assembly. 135 \textmu m is the maximum compressible structure (given the large area of our device) that can be assembled, while 35 \textmu m is the minimum structure below which the SU-8 rolls during assembly.\\
%%%%%%%%%%%%%%%%%%%%%%%%%%%%
\begin{figure}[h]
\centering
\begin{subfigure}{0.45\textwidth}
\centering
\begin{minipage}[t]{\textwidth}
\vspace{0pt}
\textbf{(a)}                                    % subfigure label
\end{minipage}
\includegraphics[width=\textwidth]{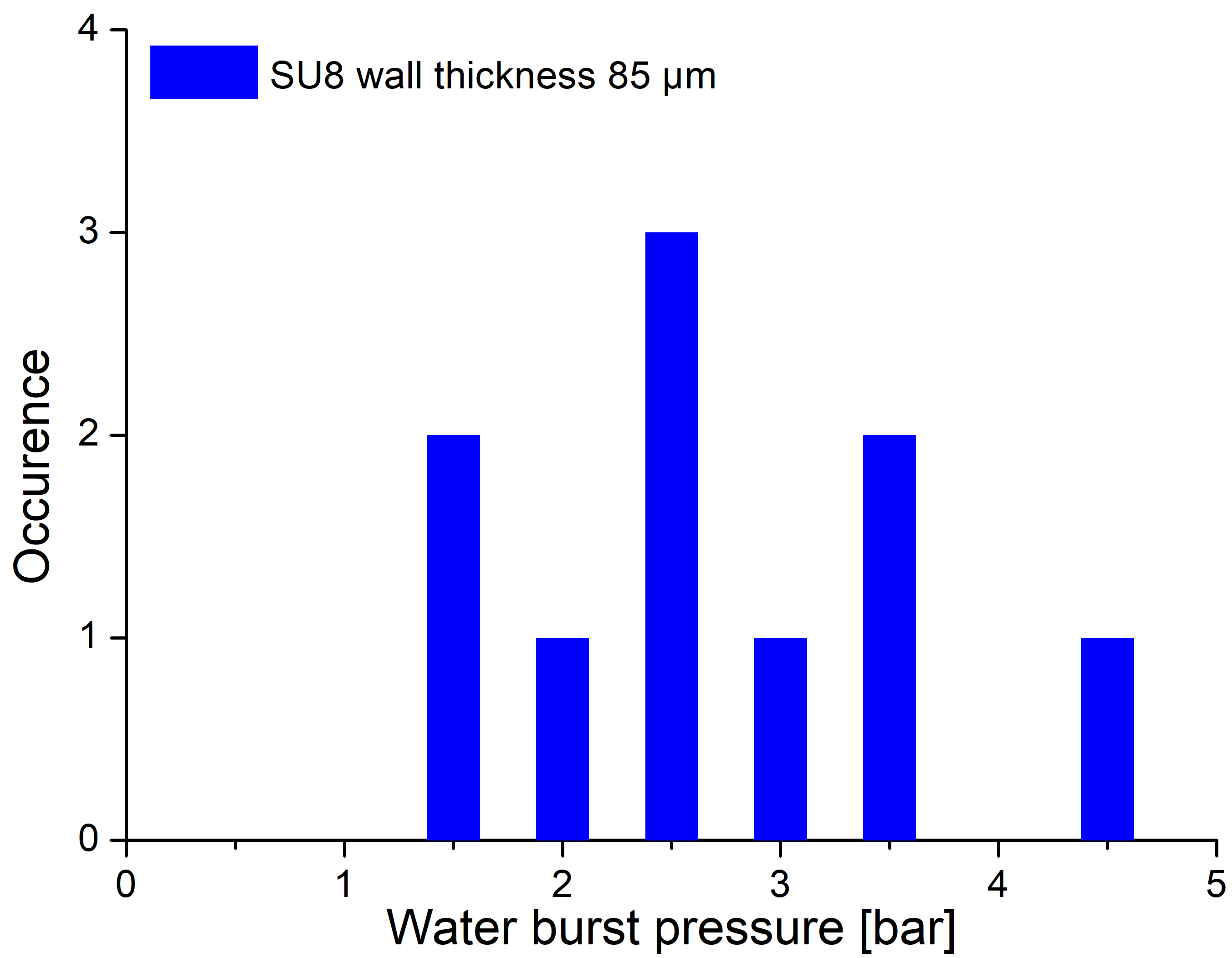} % image for subfigure (a)
\phantomcaption                                 % deactivate the caption for subfigure (a)
\label{fig: one thickness}                      % label for subfigure (a)
\end{subfigure}
\hspace{0.05\textwidth}
\begin{subfigure}{0.45\textwidth}
\centering
\begin{minipage}[t]{\textwidth}
\vspace{0pt}
\textbf{(b)} % subfigure label
\end{minipage}
\includegraphics[width=\textwidth]{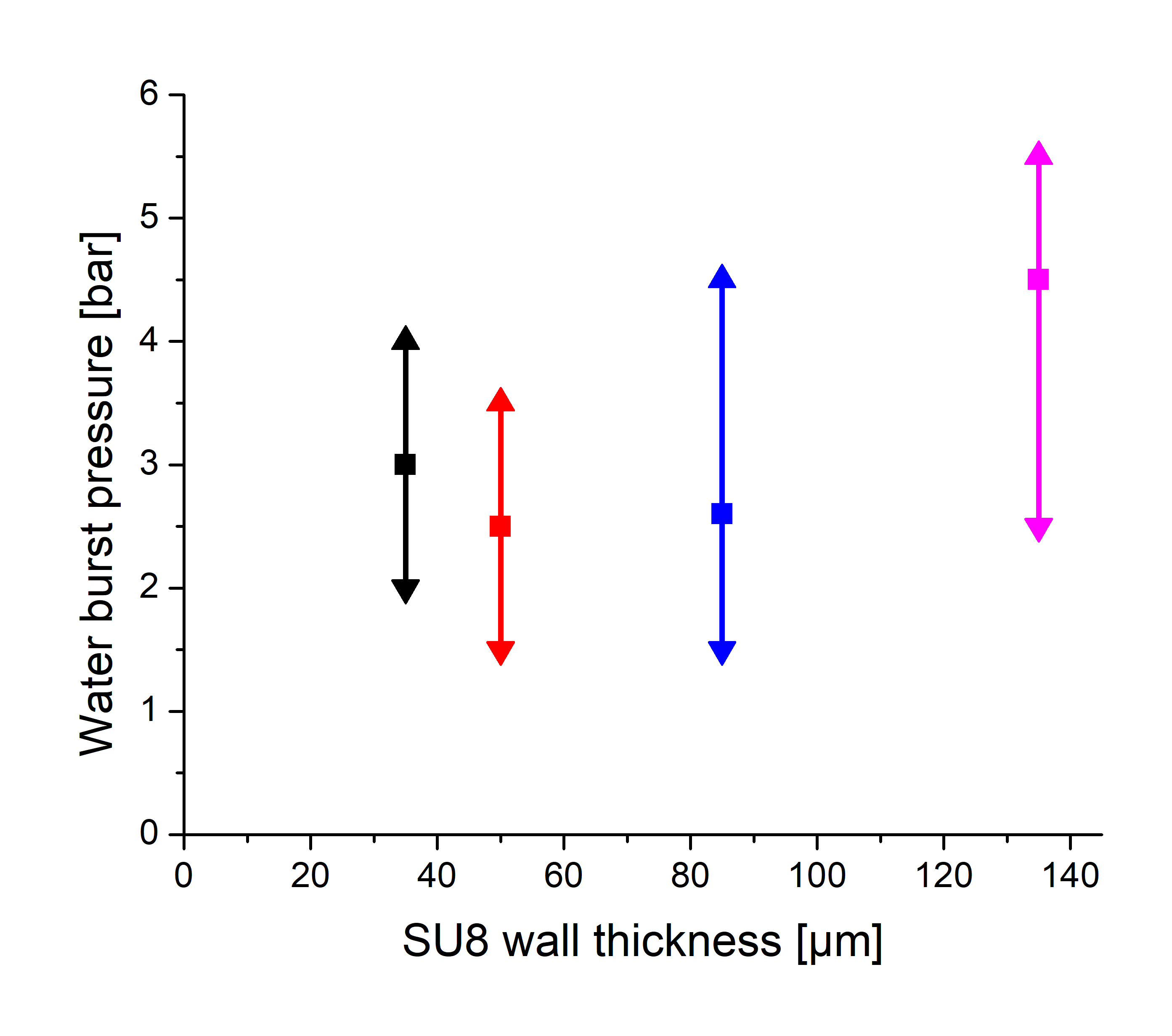} % image for subfigure (b)
\phantomcaption                               % deactivate the caption for subfigure (b)
\label{fig: all thicknesses}                      % label for subfigure (b)
\end{subfigure}
%%%%\captionsetup{type=subfigure}                 % set the caption type to subfigure
\captionsetup{font={small, it}}
\caption{Liquid water burst pressure leak-tightness. (a) Liquid water burst pressure distribution for 10 devices with 85 \textmu m thick SU-8 wall. (b) Water burst pressure distribution for various SU-8 wall thicknesses.}   % overall caption for the subfigures
\label{fig: burst pressure}                         % label for the subfigures
\end{figure}

The burst pressure distribution measured for ten devices with 85 \textmu m thick SU-8 walls is plotted in Figure \ref{fig: one thickness}. In Figure \ref{fig: all thicknesses}, we report the burst pressure distribution obtained for various values of SU-8 wall thickness. The symbols represent the mean, maximum, and minimum values of the burst pressure for each corresponding SU-8 wall thickness. The device's performance was measured for each applied pressure over 1 hour. One of the main outcomes is that the water pressure tightness does not depend on the SU-8 wall thickness. The absence of a trend of leak-tightness versus width (Figure \ref{fig: all thicknesses}) is an indication that the resistance to pressure is more governed by random imperfections of the surface and dust rather than porosities. Another interesting feature of our SU-8 devices is that they can sustain pressures ranging from 1.5 bar up to 5.5 bar. The defects during sample assembling might also explain the scattering observed in Figure \ref{fig: one thickness}.

%%%%%%%%%%%%%%%%%%%%%%%%%%%%%%%%%%%%%%%%%%%%%%
\subsection{Liquid water Porosity, Ageing, and Vacuum compatibility}
To test the liquid porosity of the SU-8 wall, we propose an original and extremely sensitive method based on the measurement of ionic conduction between two reservoirs filled with an electrolyte and separated by a SU-8 wall. Indeed, it can be assumed that if the wall exhibits percolating paths through the entire thickness then an ionic current will be observed between both reservoirs. Knowing that ionic currents as low as 1 pA were reported from extremely small porosities such as carbon nanotubes ($\sim$ 1 nm) \cite{C7NR02976D}, it can be assumed that in our case nanoporosity can be detected if a similar ionic current measurement method is used.\\
\begin{figure}[h!]
  \includegraphics[width=\linewidth]{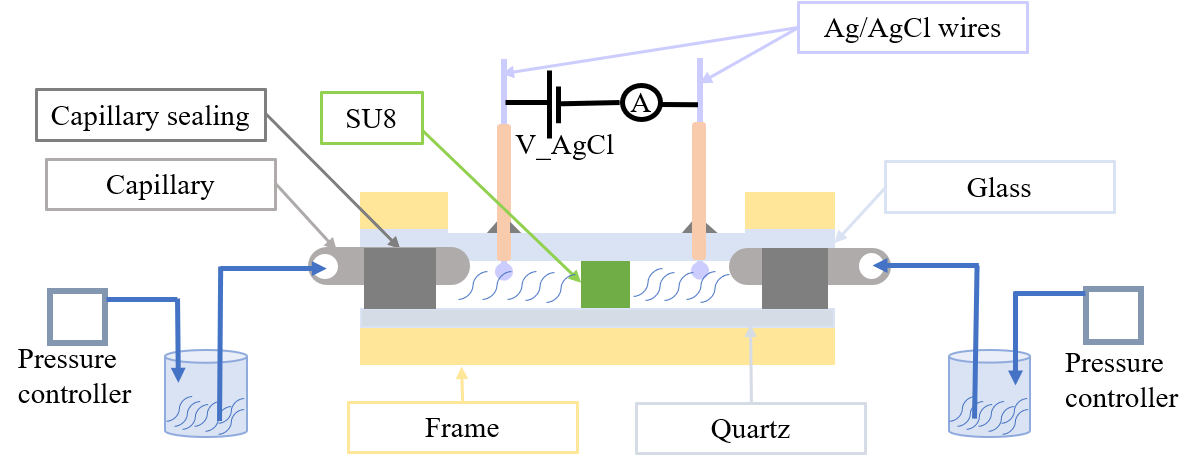}
  \captionsetup{font={small, it}}
  \caption{Schematic of the setup designed to measure the ionic conductivity between 2 micro-reservoirs filled with Milli-Q water or KCl electrolyte and separated by a SU-8 wall. }
  \label{fig:two reservoir }
\end{figure}

For that purpose, patterns of SU-8 walls were designed with 2 reservoirs on the glass substrate and assembled with an electrically insulating piece of quartz in order to lower the ionic noise \cite{citekey}. The ionic current is measured via 2 Ag/AgCl electrodes placed in each micro reservoir and tightly sealed with Stycast glue in order to avoid electrolyte leak (Figure \ref{fig:two reservoir }).\\
Finally, the electrodes are connected to a high-impedance source-amplifier (Keithley 2636B), which enables to the measurement of currents in the pA range.\\
\begin{figure}[htbp]
\centering
\begin{subfigure}[t]{0.45\textwidth}
  \centering
  \begin{minipage}[t]{\textwidth}
    \vspace{0pt}
    \textbf{(a)}
  \end{minipage}
  \includegraphics[width=\textwidth]{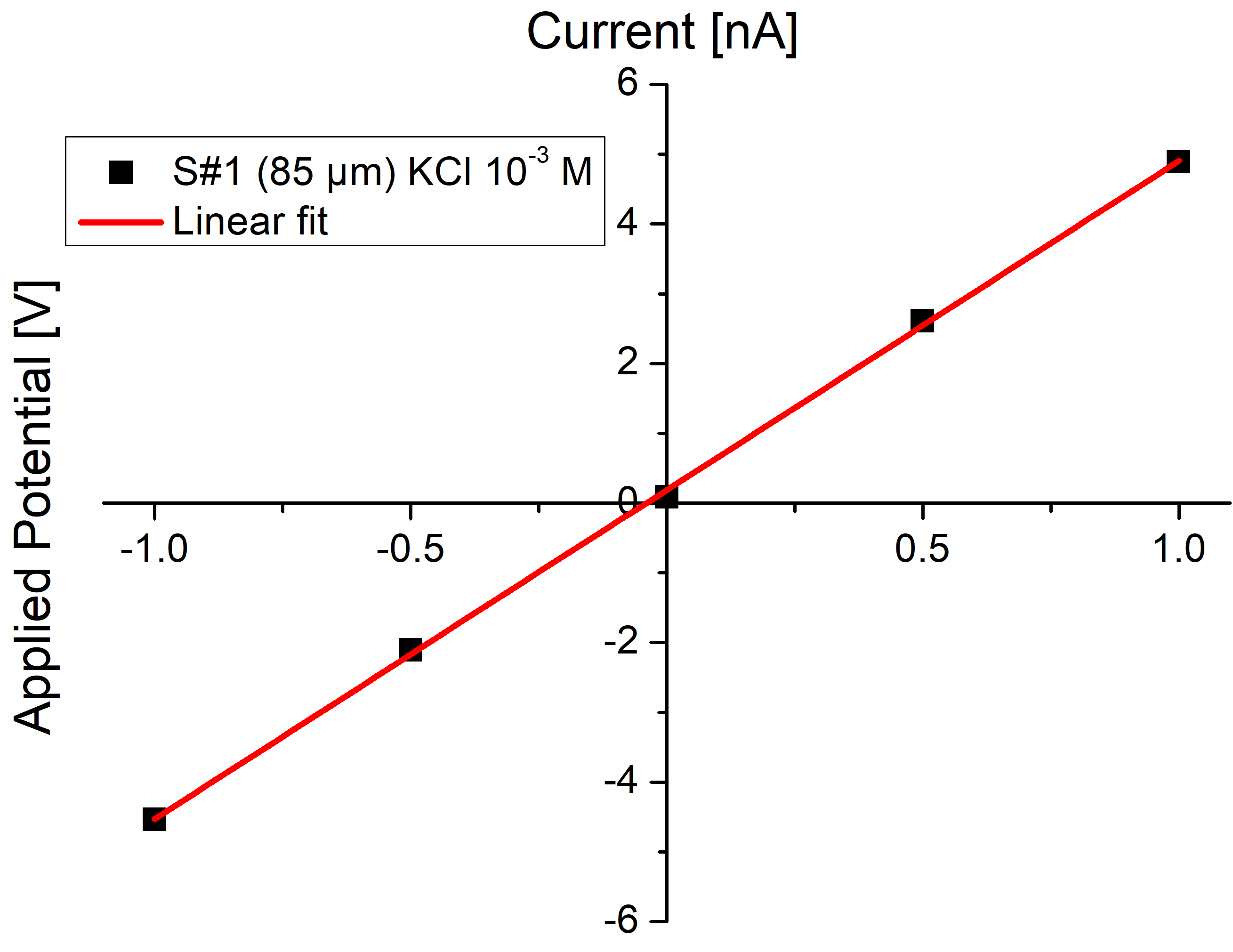}
      \phantomcaption 
    \label{fig: linear fit}
\end{subfigure}
\hspace{0.05\textwidth}
\begin{subfigure}[t]{0.45\textwidth}
  \centering
  \begin{minipage}[t]{\textwidth}
    \vspace{0pt}
    \textbf{(b)}
  \end{minipage}
  \includegraphics[width=\textwidth]{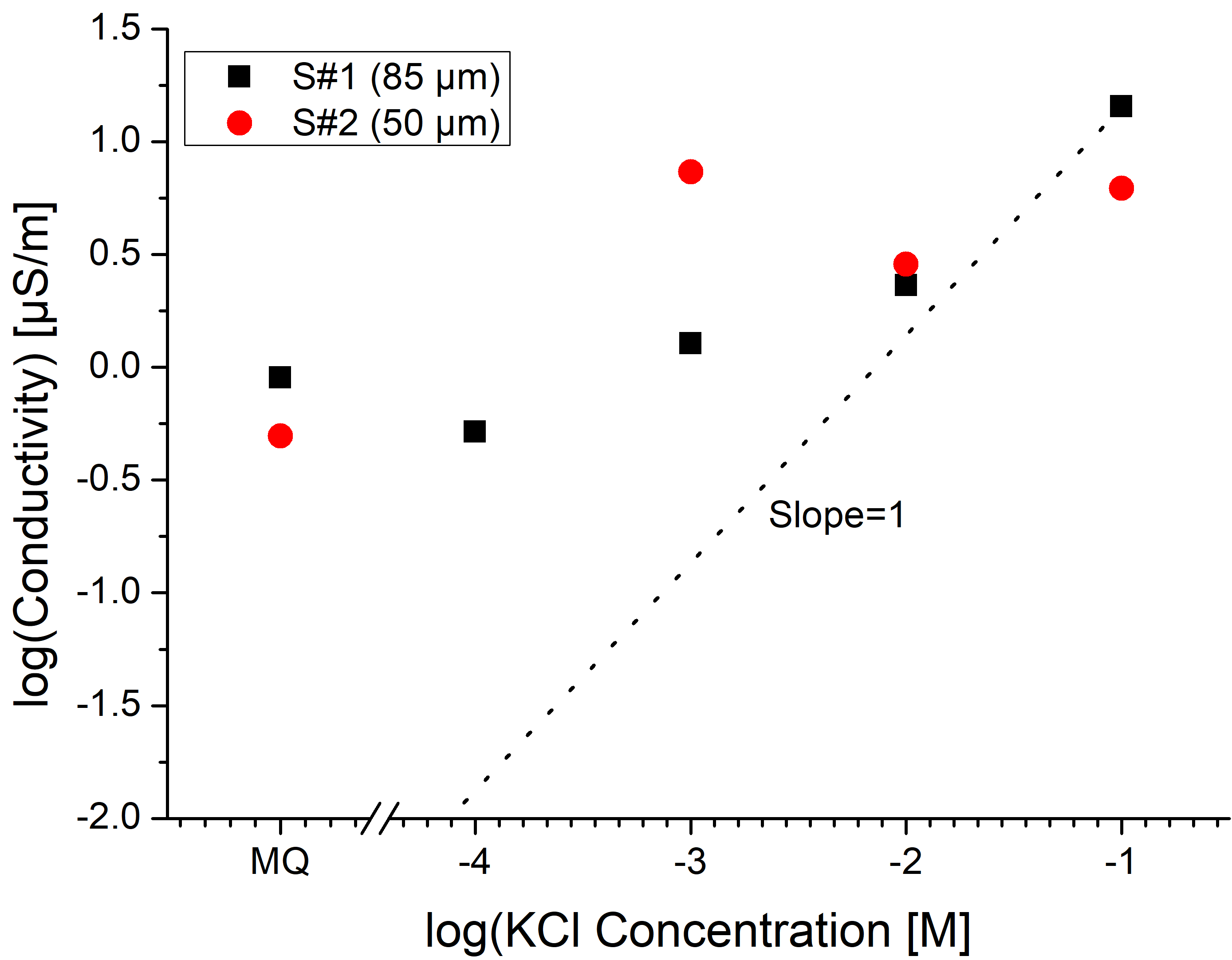}
      \phantomcaption 
    \label{fig: log scale}
\end{subfigure}
\vspace{0.05\textwidth}
\begin{subfigure}[b]{0.5\textwidth}
  \centering
  \begin{minipage}[t]{\textwidth}
    \vspace{0pt}
    \textbf{(c)}
  \end{minipage}
  \includegraphics[width=\textwidth]{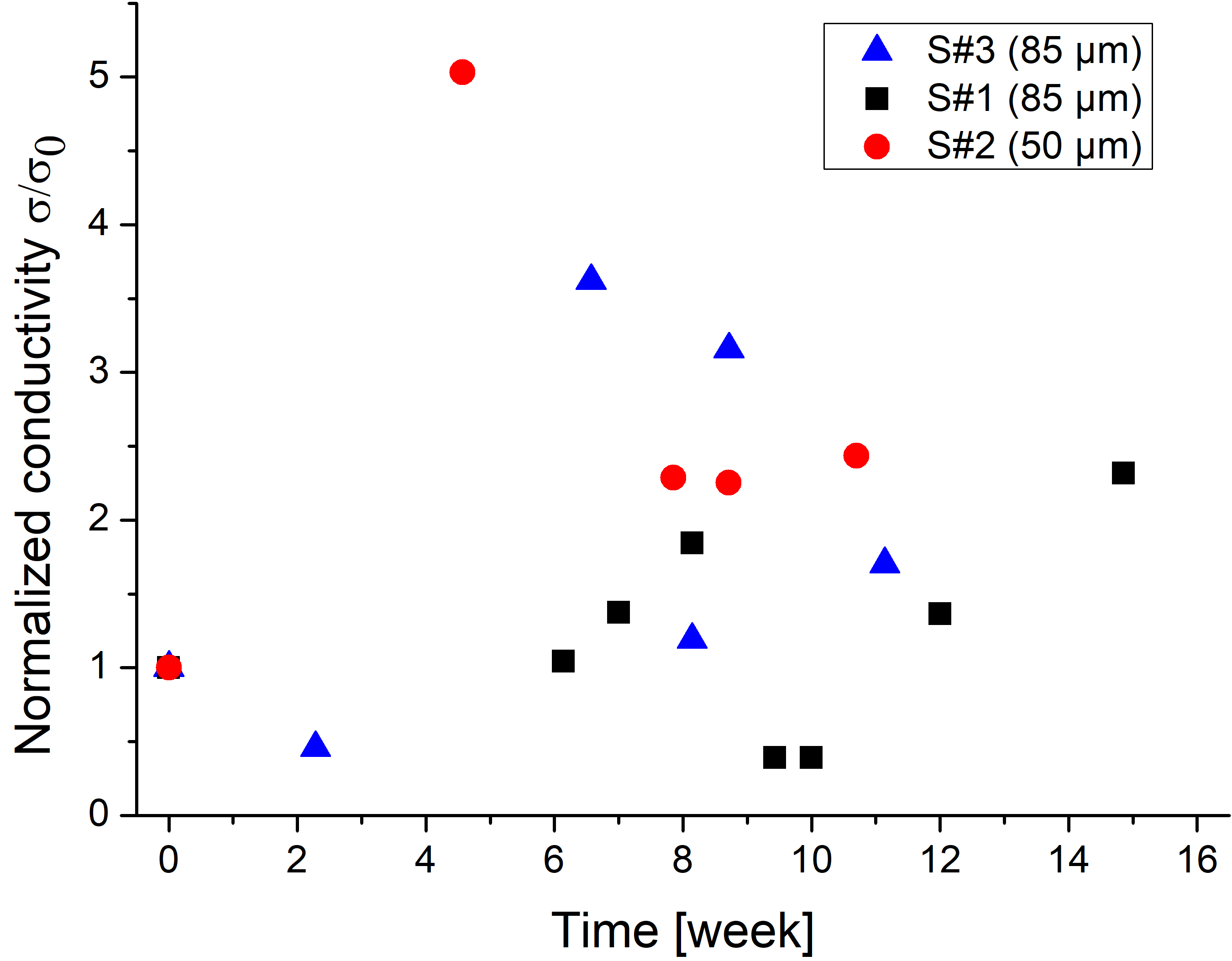}
    \phantomcaption 
    \label{fig:ageing}
\end{subfigure}
\captionsetup{font={small, it}}
\caption[]{Ionic conductivity measurement between two micro-reservoirs filled with either 300 mbar Milli-Q water or KCl electrolyte and separated by a SU-8 wall. (a) A typical example of a current/voltage plot obtained with a KCl $10^{-3}$ M electrolyte and a SU-8 wall of $w$, $l$, $h$ equal to 85 \unit{\micro m}, 10800 \unit{\micro m}, 29 \unit{\micro m}. (b) Log-log plot of the ionic conductivity versus KCl concentration for two SU-8 walls with different thicknesses. The dashed line with a slope equal to 1 represents the behavior expected for bulk ionic conductivity, i.e., Kolhrausch’s law (Equation \ref{eqn:law}). (c) Conductivity normalized by its initial value as a function of time for SU-8 walls of 3 different thicknesses (KCl concentration of 0.1 M).}
\label{fig:ionic leak tightness}
\end{figure}

First, blank measurement is carried out when both reservoirs are filled with Milli-Q water© whose resistivity is very high, about 18 M\unit{\ohm\per\cm}. Then, reservoirs are filled with KCl electrolytes at various concentrations. The resistance R of the SU8 wall is readily determined from the current/voltage plot which follows Ohm’s law, as can be seen in Figure \ref{fig: linear fit}. Then the conductivity $\sigma$ of the wall is calculated using the following equation:
%//////Reference to equation Eq.~\ref{eqn:example}
\begin{equation}
\sigma=\frac{w}{R l h} 
\label{eqn:sigma}
\end{equation}
where $w$, $l$, and $h$ are the thickness, length, and height of the SU-8 wall after baking respectively. The ionic conductivity of our set-up filled with Milli-Q water is found to be about 1 \unit{\micro\siemens\per\cm} and seems independent of the SU-8 wall thickness (Figure \ref{fig: log scale}). The dependence of the ionic conductivity with the ionic concentration in the log-log scale is reported in Figure \ref{fig: log scale}. We observe that the conductance does not follow the Kohlrausch's law for bulk conductivity of diluted electrolytes, i.e., 
\begin{equation}
\sigma=\sum_{i}\lambda_{i}C_{i} 
\label{eqn:law}
\end{equation}
where \unit{\lambda}\textsubscript{$i$} and C\textsubscript{$i$} are the molar conductivity and the concentration of the solvated ion $i$ respectively. Indeed, the log-log dependence would match with a straight line of slope 1 if Kohlrausch’s law was obeyed. The evolution observed here for device S\#1 ($w$=85 \textmu m) is similar to the usual behavior reported for ionic conductivity through nanopores, i.e., a power law dependence of the ionic conductivity with the ionic concentration at low concentration, with an exponent much lower than 1 or even as low as 0.14. We also observe a flat slope for the conductivity of S\#2 ($w$=50 \textmu m) as a function of the concentration. The fact that reducing dimension leads to the absence of transport is an indication that transport occurs through fabrication defects rather than porosities.\\
In addition, the comparison of the ionic conductivity obtained here with the one reported for similar walls made of PDMS shows that the SU-8 resin is similar or even less permeable to electrolytes and hence to water (Table S1).\\ 
The measurement of the ionic conductivity as a function of time (Figure \ref{fig:ageing}) shows that there is no significant aging effect over up to 16 weeks, while we may expect a constant decrease or increase of the ionic conductivity if the nanoporosity was closing or opening when it is filled with the electrolyte. This shows that our SU-8 wall remains watertight for a longer period of time than usually observed on micrometric walls made of other materials \cite{Klammer_2006}.\\

\begin{figure}[h!]
\centering
  \includegraphics[scale=0.7]{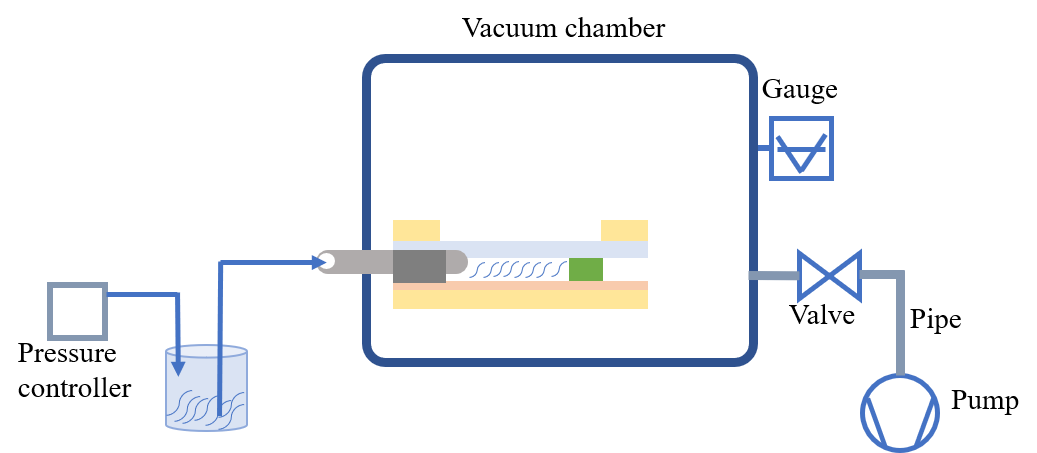}
  \captionsetup{font={small, it}}
  \caption{Schematic of the vacuum setup. The device is mounted in the chamber and SU-8 performance is tested with applying liquid water pressure under a secondary vacuum}
  \label{fig:vacuum}
\end{figure}

Finally, we assessed the ability of SU-8 to operate fluidics under a vacuum. Noteworthy, it must be recalled that curing SU-8 at 200° C for 2 hours significantly increases the reticulation rate of the resin, hence it is essential to limit outgassing as much as possible \cite{MELAI2009761}. The proposed set-up used for this test is schematically drawn in Figure \ref{fig:vacuum}. It works on the principle of the so-called pressure rising method in dynamic vacuum \cite{silverio2022overcoming}. In brief, the aim is to see an increase in the cell pressure while introducing water inside the test reservoir. The pressure rise would indicate a leak. The device as shown in Figure \ref{fig:reservoir} is mounted inside a vacuum chamber under a secondary vacuum, i.e., $5\times10^{-5}$ mbar. First, we verify that the presence of the SU-8 device does not significantly degrade the vacuum in the cell. Then, the reservoir is filled with liquid water under pressure and the change in the gauge level is recorded. It is observed that applying a pressure of 1.5 bar on water for one hour does not induce any decrease of the vacuum which remains constant at $5\times10^{-5}$ mbar under pumping. This means that the outgassing from SU-8 and the SU-8 porosity to water must be extremely small.\\

We now estimate the leak rate $Q_{leak}$ of particles, passing through our SU-8 device into the cell: $Q_{leak}$=$P_{cell}$$ \times $$D_{leak}$, with $P_{cell}$ the pressure at the level of the cell (as measured with our Gauge) and $D_{leak}$ the debit in volume at a given pressure, expressed in $L\cdot s^{-1}$.\\
Since the pressure in the cell is not changing over the leak test, one can write that the effective pumping speed at the level of the cell compensates for the leak rate of our device and the outgassing from the cell: $D_{leak}+D_{outgassing}=D_{cell}$. For the sake of simplicity, we will estimate an upper bond of $D_{leak}$ by writing:
\begin{equation}
D_{leak} \leq D_{cell}
\label{eqn:1}
\end{equation}
To compute $D_{cell}$, we recall that the flux of particles $Q$ is equal everywhere (from the cell to the pump) an can be written as:
\begin{equation}
Q=P_{cell} \times D_{cell}=P_{pump} \times D_{pump}=C(P_{cell}-P_{pump})
\label{eqn:2}
\end{equation}
Where $P_{cell}/P_{pump}$ and $D_{cell}/D_{pump}$ are respectively the pressure and the debit in volume at the level of the cell/pump, $C$ is the conductance of the circuit connecting the cell to the pump (pipes and valve). From Equation \ref{eqn:2}, we can write:
$\frac{1}{C}+\frac{1}{D_{pump}}=\frac{1}{D_{cell}}$ it comes
\begin{equation}
D_{cell}=C\frac{D_{pump}}{C+D_{pump}}
\label{eqn:3}
\end{equation}
$D_{pump}$ $\sim$ 85\,$Ls^{-1}$ is obtained from the datasheet of our pump. Because $D_{pump}$ is constant for pressures below $10^{-4}\,mbar$, we approximate that $D_{pump} (P_{pump})$ $\sim$ $D_{pump} (P_{cell})$. We will see in what follows that $D_{cell}$ is anyway dominated by $C$.\\
We now need to compute $C$, which is composed of 2 terms: $C_{pipes}$ and $C_{valve}$.
From the datasheet of our valve, $C_{valve}=160\,L\cdot s^{-1}$. $C_{pipes}$ is given in the molecular regime by \cite{agilent2022}\\
%$C_{pipes}=12.1\frac{d^3}{L}$ 
$C_{pipes}=\frac{d^3}{L}\sqrt{\frac{2 \pi k_B T}{m_{N_2}}}$, thus:\\
$C_{pipes}=16.2\,L\cdot s^{-1}$ where $L=2.4\,m$ and $d=6\,cm$ are length and diameter of the pipes respectively.\\
Finally, $\frac{1}{C}=\frac{1}{C_{pipes}}+\frac{1}{C_{valve}}$ and $C\sim$ $14.7\,L\cdot s^{-1}$.\\
Injecting in Equation \ref{eqn:3} we get $D_{cell}$ $\sim$ $12.5\,Ls^{-1}$.\\
And finally, Equation \ref{eqn:2} gives us $Q=P_{cell} \times D_{cell}$ $\sim$ $6.3\times10^{-4} \pm 1.9\times10^{-4}\, mbar\cdot L\cdot s^{-1}$, with the error bar being set by resolution of the Gauge.\\
The calculated upper bond of the leak rate is better than the leak rate corresponding to the water vapor leak tightness which is in the range of $10^{-3}\, mbar\cdot L\cdot s^{-1}$ \cite{leybold2016}.\\

\section{Conclusion}
We reported that micrometric walls made of SU-8 resin with thicknesses as low as 35 µm exhibit liquid water pressure leak tightness from 1.5 bar up to 5.5 bar and no porosity even after 2 months of aging. We also find that SU-8 can be operated under a secondary vacuum as a microfluidic seal with no degradation of the vacuum (neither outgassing of the SU-8 nor liquid water or gas permeability). Moreover, the fabrication process we propose requires neither aggressive chemical nor high temperature nor high energy plasma treatment. It thus opens a new perspective to seal microchips when sensitive surfaces or materials have to be used.

\section*{Author Contributions}
A.N, S.P., F.H. and R.L. conceived the experiment. S.P. fabricated the samples with the help of J.B, E.C., S.T., R.J, R.D., R.L. and C.R. S.P. carried out the experiment with inputs from V.J. and E.A.
A.N., F.H. and R.J supervised the work. All authors have read and agreed to the manuscript.

\section*{Conflicts of interest}
There are no conflicts of interest to declare.

\section*{Acknowledgments}
%\begin{acknowledgment}
We acknowledge financial support from CNRS-MITI (France) through a 'Momentum' grant. SP acknowledges support from ANAS (Azerbaijan) for the Ph.D. scholarship. The work was partly supported by the French Renatech network.
A CC-BY public copyright license has been applied by the authors to the present document and will be applied to all subsequent versions up to the Author Accepted Manuscript arising from this submission, in accordance with the grant’s open access conditions.
%\end{acknowledgement}
%\begin{suppinfo}
%\suppinfo{}  %{SU8 SI.pdf}

\backmatter
\bmhead{Supplementary information}

%If your article has accompanying supplementary file/s please state so here. Authors reporting data from electrophoretic gels and blots should supply the full unprocessed scans for key as part of their Supplementary information. This may be requested by the editorial team/s if it is missing. Please refer to Journal-level guidance for any specific requirements.

\bibliography{sn-bibliography}% common bib file
\end{document}